\documentclass[aps,preprint,amssymb,amsmath,nofootinbib]{revtex4}

 \newtheorem{trial definition}{Trial definition}[section]

\begin{document}
 \title{Two spaces that already found their geometer in the thirties}
 \author{Gavriel Segre}
 \homepage{http://www.gavrielsegre.com}
 \email{info@gavrielsegre.com}
 \bigskip
 \begin{abstract}
 Giorgio Parisi's  recent speculations on the concept of continuous
 dimension are compared with Von Neumann's serious work
 \end{abstract}
 \maketitle
 \newpage
 On a paper appeared recentely at the Arxiv \cite{Parisi-02}
 Giorgio Parisi expresses his opinion that some geometer has to
 mathematically formalize the concepts of
 non-integer-dimensional vector spaces and of matrix-action on
 them, euristically occurring many times in Physics, e.g. in the dimensional regularization usual in Quantum Field Theory
 \cite{Weinberg-96} or in the $ \epsilon$-expansion of Statistical
 Mechanics \cite{Fisher-99} and claimed to be important for Spin
 Glass Theory (perhaps also as to the serious literature
 \cite{Bolthausen-02}).

 As any serious physicist, mathematician or mathematical-physicist should know such a geometer has a name:
 John Von Neumann, who realized what Parisi is looking for yet in
 the thirties, in a way I will briefly review demanding to some
 serious literature for details
 \cite{Von-Neumann-88},
 \cite{Petz-Redei-95},\cite{Redei-98},\cite{Redei-01}, \cite{Connes-94}
 \cite{Connes-98},\cite{Landi-97}, \cite{Gracia-Bondia-Varilly-Figueroa-01}, \cite{Segre-02}.

 Given a D-dimensional linear space V on the field $ {\mathbb{K}}
 $  let us define its \textbf{projective geometry} $ PG(V) $ as the set of
 all its linear subspaces:
\begin{equation}\label{eq:projective geometry of a linear space}
  PG(V) \; := \; \bigcup_{k=0}^{D} G_{k,D}({\mathbb{K}})
\end{equation}
where $  G_{k,D}(V) $ is the $k^{th}$ Grassmannian of V, namely
the set of all the k-dimensional linear subspaces of V.

Introduced on $ PG(V) $ the partial ordering relation:
\begin{equation}\label{eq:partial ordering in a projective geometry}
  a \, \preceq \, b \; := \; a \, \subseteq \, b  \; \; a \, , \,
  b \in  PG(V)
\end{equation}
the partially-ordered-set $ ( PG(V) \, , \, \preceq ) $ is an
\textbf{atomic lattice}, with:
\begin{align}
  a \wedge b & \, := \,  a \bigcap b \\
  a \vee b & \, := \,  a \bigoplus b
\end{align}
whose atoms are the one-dimensional subspaces, namely the
elements of $ G_{1,D}({\mathbb{K}}) $, said the \textbf{points}
of the \textbf{projective geometry} of V.

It may be easily verified that the following map:
\begin{equation}\label{eq:dimension function}
  d(a) \; := \; dim(a) \; \; a \in PG(V)
\end{equation}
is a \textbf{dimension function} for the lattice $ ( PG(V) \, ,
\, \preceq ) $. Since:
\begin{equation} \label{eq:atomic modular projective geometry}
  \infty \; \notin \; Range(d) \; = \; \{ 0 \, , \, 1 \, , \,
  \cdots \, , \, D \}
\end{equation}
the lattice  $ ( PG(V) \, , \, \preceq ) $ admits a finite
dimension function and is, conseguentially, \textbf{modular}.

Indeed  eq.\ref{eq:atomic modular projective geometry} completely
characterizes the nature of the analyzed projective geometry:
\begin{itemize}
  \item  the discrete nature of Range(d), encoding the \textbf{atomicity} of PG(V),
tells us that that it doesn't admit "intermediate subspaces"
lying between points and lines, between lines and planes an so on
  \item the finite nature of Range(d), encoding the
  \textbf{modularity} of  PG(V), tells us that that it doesn't
  admit subspaces of arbitrary large dimensionality
\end{itemize}

\smallskip

Let us now suppose that V is an infinite-dimensional, separable
vector space.

$ ( PG(V) \, , \, \preceq ) $ is again a lattice on which the map
d defined in eq.\ref{eq:dimension function} is again a dimension
function. In this case, anyway, such a map is no more finite:
\begin{equation} \label{eq:atomic nonmodular dimension function}
  \infty \; \in \;  Range(d) \; = \;  \{ 0 \, , \, 1 \, , \,
  \cdots \, , \, \infty \}
\end{equation}
and the lattice $ ( PG(V) \, , \, \preceq ) $ is no more modular.

Indeed, again, eq.\ref{eq:atomic nonmodular dimension function}
completely characterizes the nature of the analyzed projective
geometry:
\begin{itemize}
  \item  the discrete nature of Range(d), encoding the \textbf{atomicity} of PG(V),
tells us that that it doesn't admit "intermediate subspaces"
lying between points and lines, between lines and planes an so on
  \item the nonfinite nature of Range(d), encoding the
  \textbf{nonmodularity} of  PG(V), tells us that it
  admits subspaces of arbitrary large dimensionality
\end{itemize}

\smallskip

Let us observe, at this point, that in both the analyzed cases
the structure of the projective geometry PG(V) rules the
Representation Theory of Lie groups on G, as may be easily
understood looking at reduction of representations.

\bigskip

The correct way of introducing "intermediate subspaces" of
suitable projective geometries lied at the heart of Von Neumann's
classification of factors (completed by Alain Connes):

the key ingredient is the notion of \textbf{relative dimension}:

given a  \textbf{noncommutative space}, namely a Von Neumann
algebra A acting on an Hilbert space $ {\mathcal{H}} $, two
projectors $ p_{1} $ and $ p_{2} $ belonging to A are called
\textbf{equivalent w.r.t. A} (this fact being denoted as $ p_{1}
\, \sim_{A} \, p_{2} $) if there exist in A a partial isometry
between $ Range(p_{1}) $ and  $ Range(p_{2}) $, a condition that
may be intuitively seen as the requirement that, \textbf{"from the
viewpoint of A"}, the dimensionality of the spaces on which $
p_{1} $ and $ p_{2} $ project are equal.

Such a notion of \textbf{relative equivalence} of projections
immediately induces a partial ordering $ \leq_{A} $ on the
projections of A, according to which the projection  $ p_{1} $ is
considered less or equal to another projection $ p_{2} $
\textbf{w.r.t. A} if is has a \textbf{dimension w.r.t. A} less or
equal to that of $ p_{2} $.

The resulting notion of \textbf{relative dimension} $ d_{A} $
w.r.t. a noncommutative space A allowed Von Neumann to introduce
his celebrated classification of factors:
\begin{itemize}
  \item  A  is of \textbf{type n-dimensional, discrete} ( $ Type(A ) \, = \, I_{n} $
  ) iff $ Range ( d_{A} ) \, = \, \{ 0 , 1 , \cdots \, n \} $
  \item A  is of \textbf{type infinite, discrete} ( $ Type(A ) \, = \, I_{\infty} $
  ) iff $ Range ( d_{A} ) \, = \, \{ 0 , 1 , \cdots \, \infty \} $
  \item A  is of \textbf{type finite, continuous} ( $ Type(A ) \, = \, II_{1} $
  ) $ Range ( d_{A} ) \, = \, [ 0 \, , \, 1 ] $
  \item A  is of \textbf{type infinite, continuous} ( $ Type(A ) \, = \, II_{\infty} $
  ) iff $ Range ( d_{A} ) \, = \, [ 0 \, , \, \infty ] $
  \item A  is of \textbf{type purely infinite} ( $ Type(A ) \, = \, III $
  ) iff $ Range ( d_{A} ) \, = \, \{ 0 \, , \, \infty \} $
\end{itemize}
As it has been strongly emphasized by Miklos Redei, Von Neumann
explicitely realized that what he was doing was nothing but
\textbf{Theory of Noncommutative Cardinality}, i.e. the theory of
\textbf{noncommutative cardinal numbers} describing the "sizes"
(and  "infinity's degrees")  of \textbf{noncommutative sets}:
\begin{center}
   \textit{"$ \cdots $ the whole algorithm of Cantor theory is such that the most of it
   goes over in this case. One can prove various theorems on the additivity of equivalence and the
   transitivity of equivalence, which one would normally expect, so that one can introduce a theory of alephs here, just as in set
   theory. $ \cdots $ I may call this dimension since for all matrices of the ordinary space, is nothing else but dimension"
   (Unpublished, cited in \cite{Redei-98})}

   \textit{"One can prove most of the Cantoreal properties of finite and
   infinite, and, finally, one can prove that given a Hilbert space and a ring in it , a simple ring in it,
   either all linear sets except the null sets are infinite (in which case this concept of alephs gives you
   nothing new), or else the dimensions, the equivalence classes, behave exactly like numbers
   and there are two qualitatively different cases. The dimensions either behave like integers, or else
   they behave like all real numbers. There are two subcases, namely there is either a finite top or
   there is not"
   (Unpublished, cited in \cite{Redei-98})}
\end{center}
This fact has induced me \cite{Segre-02} to suggest a notation
remarking it by the explicit introduction of the following notion
of noncommutative cardinality:
\begin{equation*}
  card_{NC} (A) \; := \; \int_{{\mathcal{Z}}(A)}^{\otimes} card_{NC} ( A_{\lambda} ) \, d \nu (
  \lambda )
\end{equation*}
where:
\begin{equation*} \label{eq:factor decomposition of a Von Neumann algebra}
  A \; = \; \int_{{\mathcal{Z}}(A)}^{\otimes} A_{\lambda} \, d \nu ( \lambda
  )  \; , \;
    {\mathcal{Z}}(A_{\lambda}) \; = \; \{ {\mathbb{C}} \, {\mathbb{I}} \} \; \; \forall \lambda \in  {\mathcal{Z}}(A)
\end{equation*}
and:
\begin{itemize}
  \item A HAS NONCOMMUTATIVE CARDINALITY EQUAL TO $ n \in {\mathbb{N}}$:
\begin{equation}
  cardinality_{NC}(A) \, = \, n \; := \; Type(A) \, = \, I_{n}
\end{equation}
 \item A HAS NONCOMMUTATIVE CARDINALITY EQUAL TO $ \aleph_{0} $:
\begin{equation}
  cardinality_{NC}(A) \, = \,\aleph_{0}   \; := \; Type(A) \, = \, I_{\infty}
\end{equation}
 \item A HAS NONCOMMUTATIVE CARDINALITY EQUAL TO $ \aleph_{1} $:
\begin{equation}
  cardinality_{NC}(A) \, = \,\aleph_{1}   \; := \; Type(A) \, \in \,
  \{ II_{1} , II_{\infty} \}
\end{equation}
 \item A HAS NONCOMMUTATIVE CARDINALITY EQUAL TO $ \aleph_{2} $:
\begin{equation}
  cardinality_{NC}(A) \, = \,\aleph_{2}   \; := \; Type(A) \, = \,
  III
\end{equation}
\end{itemize}
Let us consider now a noncommutative space A of noncommutative
cardinality $ \aleph_{1} $:

all the category-equivalence's theorems giving foundations to
Noncommutative Geometry   allow to look at the lattice of
projections $ P(A) $  as the \textbf{projective geometry} of A.

$ P(A) $ is, anyway, \textbf{nonatomic}, i.e. it is a projective
geometry \textbf{without points}, explicitely manifesting the
phenomenon of \textbf{continuous geometry}: according to the basic
"noncommutative metaphore" (formalized by the mentioned
category-equivalence's theorems) looking at A "as if it was an
algebra of functions on a new kind of space"  we can infer that,
in this case "such a strange kind of space " has subspaces of any
integer dimension $ \epsilon $ between zero and one.

If A is finite, furthermore, the geometric structure of these
"strange kind of subspaces of noninteger dimension " is
astonishing: considered a subfactor B of A the ratio between the
\textbf{relative dimension w.r.t. B } and the \textbf{relative
dimension w.r.t. A} allows to define the \textbf{Jones' index of
B w.r.t. A} whose role in Knot Theory realizes a link to Algebraic
Topology that has led to some of the more extraordinary results
of the last decade both in Mathematics  and in Physics
\cite{Birman-97}, \cite{Murasugi-96}, \cite{Kodiyalam-Sunder-01},
\cite{Ohtsuki-02}.

\smallskip

 As to the second mathematical object Parisi is looking for, namely a mathematical
formalization of the action of matrices on noninteger-dimensional
linear spaces, let us observe that the structure of the
projective geometry PG(A) rules the Representation Theory of G by
automorphisms.

\smallskip

The usual application of Noncommutative Geometry as to Serious
Quantum Statistical Mechanics and Serious Quantum Ergodic Theory
 \cite{Thirring-81}, \cite{Thirring-83}, \cite{Simon-93},
\cite{Ruelle-99}, \cite{Benatti-93}, \cite{Alicki-Fannes-01}
could lead to think  erroneously that the only role Noncommutative
Geometry plays in Physics concerns its role as to Quantum
Mechanics:

this is not, anyway, true since Noncommutative Geometry may be
used also to analyze many mathematical structures appearing in
Classical (i.e. non quantum) Physics, among which it has
certainly to be mentioned the other great mathematical adventure
concerning spaces with non-integer dimension: Fractal Geometry
\cite{Falconer-90}:

as an example, introduced the  spectral triple $ ( A \, , \,
{\mathcal{H}} \, , \, D ) $ of 1-dimensional noncommutative
calculus:
\begin{itemize}
  \item
\begin{equation}
  {\mathcal{H}} \; := \; L^{2} ( S^{(1)} , d \vec{x}_{Lebesgue} )
\end{equation}
  \item
\begin{equation}
  A \; := \; L^{\infty} ( S^{(1)} , d \vec{x}_{Lebesgue} )
\end{equation}
where a function $ f \in A $ is seen as a multiplication operator:
\begin{equation}
  ( f \psi ) (t) \; := \; f(t) \psi (t) \; \; f \in A , \psi \in {\mathcal{H}}
\end{equation}
  \item D is the linear operator on $ {\mathcal{H}} $ defined by:
\begin{equation}
  D e_{n} \; := \; sign(n) e_{n} \: , \: e_{n} ( \theta ) \; := e^{ i n \theta
  }  \; \forall \theta \in S^{(1)}
\end{equation}
\end{itemize}
and denoted by D the Hausdorff dimension of the Julia set $ J [
p_{c} (z) ] $:
\begin{equation}
  J [ p_{c} (z) ] \; = \; \partial \, \{ z \in {\mathbb{C}} : \sup_{n \in
  {\mathbb{N}}} | p_{c}^{(n)} (z)| < \infty \}
\end{equation}
of the quadratic maps on the complex plane $ p_{c} (z) \, := \,
z^{2} + c $, Alain Connes proved that:
\begin{enumerate}
  \item $ | d_{NC} Z | $ is an infinitesimal of order $
  \frac{1}{D} $
  \item
\begin{equation}\label{eq:integration of continuous functions on Julia's
set}
  \exists \lambda > 0 \; : \; ( \int_{J [ p_{c} (z) ]} f d \Lambda_{D} ) \, = \;
  \lambda \int_{NC} f(Z) | D |^{ - 1}  | d_{NC} Z |^{D} \; \; \forall f \in C( J [ p_{c} (z)
  ] )
\end{equation}
where $ d \Lambda_{D} $ is the Hausdorff measure on $ J [ p_{c}
(z) ] $.
\end{enumerate}
The eq.\ref{eq:integration of continuous functions on Julia's
set} tells us that the integral w.r.t. the Hausdorff measure of
continuous functions over the Julia set $ J [ p_{c} (z) ] $ may
be computed as a noncommutative integral in the spectral triple $
( A \, , \, {\mathcal{H}} \, , \, D ) $.

Since the Mandelbrot's set $ {\mathcal{M}} $  is linked to the
family of Julia sets $ J [ p_{c} (z) ] $ by the condition:
\begin{equation}
  {\mathcal{M}} \; = \; \{ c \in {\mathbb{C}} \,  : \,  J [ p_{c} (z) ]
  \text{ is connected } \}
\end{equation}
eq.\ref{eq:integration of continuous functions on Julia's set}
could be useful to investigate some of the still unknown
properties of one of the most precious diamonds of Fractal
Geometry: Mandelbrot's set $ {\mathcal{M}} $.

\smallskip

Beside all the mentioned reasons to think that the geometer
Parisi is looking for was John Von Neumann in the thirties
(together with his most prominent successor Alain Connes), let us
observe that the notion of non-integer dimensional vector space
obtained through Noncommutative Geometry strongly differs from
the strategy Parisi himself proposes in order of obtaining a
definition of a non-integral dimensional vector space, strategy
consisting in:
\begin{enumerate}
  \item considering a suitable  family $ \{ S_{i} \} $ of
  submanifolds of the D-dimensional euclidean space $ ( {\mathbb{R}}^{D} \, , \, \delta = \delta_{\mu \nu } dx^{\mu} \bigotimes dx^{\nu} ) $
  \item making the analitical continuation in D of the formulas:
\begin{equation}
  \mu(S_{i}) \; = \, f_{i} (D)
\end{equation}
expressing the (induced)-measure of each submanifold $ S_{i} $.

\end{enumerate}
It is sufficient, anyway, to think just some minute to realize how
sick is Parisi's illusion that such a strategy could lead to
characterize in a meaningful way a notion of non-integer
dimensional linear space:

which is the  family $ \{ S_{i} \} $ to be considered ?

and why, even before of taking any analytical continuation in D,
one should think that their  measures encode all the geometrical
structure of $ ( {\mathbb{R}}^{D} \, , \, \delta = \delta_{\mu
\nu } dx^{\mu} \bigotimes dx^{\mu} ) $ such as, for example, the
measure of a submanifold not belonging to the chosen family?

\end{document}